\documentclass[final,3p,times]{elsarticle}

\usepackage{epsfig}

\usepackage{amssymb}
\usepackage{amsmath}

\usepackage[T1]{fontenc}

\journal{Physica A}

\begin{document}
	
	\begin{frontmatter}
		
		
		
		\title{Diffusion on hierarchical systems of weakly-coupled networks}
		
		
		\author[WUT]{Grzegorz Siudem}
		\author[WUT,NIAS,ITMO]{Janusz A. Ho{\l}yst}
		
		\address[WUT]{Faculty of Physics, Center of Excellence for Complex Systems Research, Warsaw University of Technology, Koszykowa 75, PL-00-662 Warsaw, Poland}
		
		\address[NIAS]{Netherlands Institute for Advanced Study in the Humanities and Social Sciences, 
			Meijboomlaan 1, 2242 PR Wassenaar, The Netherlands}
		
		\address[ITMO]{ITMO University, 49 Kronverkskiy av., 197101 Saint Petersburg, Russia}

		\begin{abstract}
			
			We analyzed diffusion dynamics on weakly-coupled networks (interconnected networks) by means of separation of time scales. Using an adiabatic approximation we reduced the system dynamics to a Markov chain with aggregated variables and derived a transport equation that is analogous to Fick's First Law and includes a driving force. Entropy production is a sum of microscopic entropy transport, which results from the particle's migration between networks of different topologies and macroscopic entropy production of the Markov chain. Equilibrium particles partition between different sub-networks depends only on internal sub-network parameters.
			By changing structure of  networks  one can not only modify diffusion constants but can also induce or  reverse  the direction of the particles' flow between different networks. Our framework, confirmed by numerical simulations, is also useful for considering diffusion in nested systems corresponding to  hierarchical networks with  several different time scales thus it can serve to uncover hidden hierarchy levels from observations of diffusion processes.
			
		\end{abstract}
		
		\begin{keyword}
			complex networks \sep hierarchical systems \sep diffusion dynamics
			\PACS 05.10.-a \sep 89.75.-k
			
		\end{keyword}
		
	\end{frontmatter}
	
	

	\section{Introduction}
	The main motivation behind our work was the question 'how does   the random walk spread particles between weakly coupled networks?'. This idea belongs to intensively investigated topics related to diffusion processes taking place in various complex systems including  hierarchical, coupled or multileveled networks  (see e.g.   \cite{Arenas,Nemarzadeh,kivela}, we present a  longer list of the references below).  Surprisingly when we finished our work we realized that this kind of approach can be also used as a tool for the community  or hierarchy detection  in real or artificial networks (see Fortunato's user guide \cite{Fortunato2016} and references below). 
	
	Starting from the seminal work of Albert and Barab\'{a}si \cite{Barabasi} a lot  of attention was paid to  dynamical processes  taking place in the systems of complex networks. Examples are investigations of phase transition  of Ising model  at scale-free  Barab\'{a}si  Albert graph \cite{Aleksiejuk,Bianconi}  or models of epidemical processes at various complex networks, see e.g. \cite{Pastor-Satorras,Moreno,Cha,Doerr}.

	Many natural and techno-social systems can be described as coupled  networks or interconnected networks  and they belong to a larger class of so-called multilayer networks \cite{kivela,boccaletti,Domen2,kurant}.  In fact, real networks (natural or artificial) are rarely  single  structureless systems  and they tend to consist of sets of sub-networks with dense internal connections (e.g. network communities, see \cite{Girvan, Newman,Eriksen,simonsen} and sparse inter-network links  \cite{buldyrev,gu,Halu,suchecki1,suchecki2} .
	
	In this paper we consider the fundamental issue of diffusion taking place in  weakly coupled  networks that form nested hierarchical structures \cite{hierach}. Diffusion processes on complex networks were widely investigated, e.g. \cite{Eriksen,simonsen,noh,gomez,sinatra,meyer,hwang, gomez2,Rosvall,Rosvall_Hierarch,Czaplicka1,Czaplicka2}, usually by transferring microscopic particle density dynamics to Markov Chains \cite{cover,norris}. The ergodic distribution of diffusing particles focuses on nodes with high random-walk centrality parameter values \cite{Eriksen, simonsen,noh}. If the diffusion takes place on a multilayer network with different diffusion constants at every layer \cite{gomez}, the effective diffusion time scale is shorter than the corresponding time scale calculated for the slower independent network. 
	
	Another class of problems related to our approach is detection of networks communities  by considering various properties of an auxiliary  diffusion process on networks.  Eriksen et al.  \cite{Eriksen,simonsen}  showed that spectral properties of diffusion matrix (e.g. components of slowest decaying eigenmodes) can be successfully used to uncover hidden network modules. Girvan and  Newman
	\cite{Girvan,Newman} proposed a new class  of algorithms  of  community detection  based on recursive removing of links possessing    highest {\it link betweenness}.
	Rosvall and Bergstrom \cite{Rosvall} applied the Huffman code \cite{Huffman} to find the optimal community partition by minimizing the average description length of diffusive paths that were represented  as appropriate binary strings.  
	
	The presented in our paper approach is similar to the methodology proposed  by Liu and Liu  \cite{Liu}, where hierarchical community structures were detected  using  a {\it coarse-grained diffusion}. Here we  consider networks with known modular topology  and we  reconstruct   coarse-grained  diffusion dynamics at different aggregation levels.  Similarities between both approaches appear when comparing, for example, the coarse-grained particles' distributions at the community level (cf. e.g. Eq. (\ref{Nainfty}) in this paper and Eq. (11) in \cite{Liu}) as well as corresponding Markov transition matrices.
	
	On the other hand  our approach of time scale separations  is  more general  than the previous studies since it allows to take into account additional crucial network attributes, i.e. {\it fitness parameters} of individual nodes. The idea of fitness parameters (sometimes called hidden variables) was previously intensively applied for a wide class of mathematical models for complex systems \cite{Bianconi2001,Caldarelli2002,Soderberg,Boguna,Fortunato,Hoppe}. Those fitness parameters affect to models for evolution of scale-free networks \cite{Bianconi2001, Caldarelli2002}, inhomogeneous sparse random graphs \cite{Soderberg}, or  correlated random networks \cite{Boguna}. 
	
	The remainder  of this paper is organized as follows. In the next section  a model of particles diffusing on coupled networks with fitness factors is introduced.  In  Sec.~\ref{SecIII}  we  present our analytical approach for coarse-grained diffusion on two weakly coupled networks  that bases upon a time-scales separation and an adiabatic approximation. Results of numerical simulations of particles diffusing in coupled  networks and estimated values of inter-network diffusion constants are presented in  Sec. \ref{SecIV}. Sec. \ref{Sec_postrec} is devoted to  detection of direction of diffusion flux  and  Sec. \ref{SecV} to the entropy production. Sec. \ref{SecVI} contains more general results for  diffusion on many coupled networks and diffusion on hierarchical systems of weakly coupled networks with a nested topology is investigated in Sec. \ref{SecVII}. The paper concludes with a discussion of  main results (Sec. \ref{SecVIII}). Appendix includes  an extension of our formalism to weighted coupled networks with self-loops (\ref{sec:non}).
	
	\section{Model definition} \label{SecII}
	Here we   introduce basic concepts of our framework that allow  to study  the diffusion process on  {\it weakly connected} sub-networks with fitness functions (for brevity we will omit  the prefix {\it sub} when it does not lead to confusion). The assumed weak inter-network coupling permits  us to distinguish a fast diffusion process inside every sub-network from a slower diffusion process between different sub-networks and, eventually, to find an approximated analytical solution for this problem in an adiabatic approximation. For brevity in this Section  we demonstrate our framework for the simplest case of a pair of connected sub-networks  \cite{buldyrev, suchecki2, shao, gao, gao2,gao3}. Later  in Sec. \ref{SecVI} we extend the methodology to hierarchical structures with a nested topology of weakly interacting sub-networks  \cite{hierach,Ravasz,Sales-Pardo,Clauset}.
	
	We consider diffusion as a pure random walk in graphs, so it is natural that the problem's natural language are Markov processes \cite{norris}. In the following paragraphs we introduce the form of the Markov's matrix, see Eq. (\ref{eq:dynamics}), which governs the particles dynamics. Using the Markov processes formalism one gets the equilibrium density of the process, see Eq. (\ref{eq:n_i}) and the corresponding entropy -- Eq. (\ref{eq:S1}). Then we consider a specific case of two coupled networks, when  the adjacency matrix takes the form given by Eq. (\ref{eq:matrix}) and in view of the complexity of the general form of Eq. (\ref{eq:dynamics})  in Sec. \ref{SecIII} we introduce  a coarse-grained approach for the problem.
	
	Imagine a situation where $N$ non-interacting particles randomly diffuse within an un-directed and un-weighted network $\mathcal{A}$ with an adjacency matrix $A\in\mathbb{M}^{M\times M}(\{0,\,1\})$. Each node $i$ ($i=1,\,2,\,\dots,\, M$) possesses a specific time-independent  attribute  called fitness (or attractiveness) $f_i>0$ that determines how strongly the node can attract moving particles. Changes of particle's density $n_i(t)$ at a node $i$ are given by
	\begin{equation}\label{eq:dynamics}
	n_i(t+1)=f_i\sum_{k=1}^M\frac{A_{ik}}{g_k}n_k(t),
	\end{equation}
	where $g_k=\sum_l A_{lk}f_l$ is equal to the total attractiveness of the $k$-th node neighborhood.  
	Equilibrium density of particles in the node $i$ follows from an ergodic distribution $\mu_i$ of a corresponding Markov chain \cite{norris} 
	\begin{equation}\label{eq:n_i}
	\mu_i=\frac{f_ig_i}{\sum_{k=1}^{M}f_kg_k},
	\end{equation}
	as $n_i= N \mu_i$. The ergodic solution (\ref{eq:n_i}) exists if the graph $\mathcal{A}$ is connected and it contains at least one  odd cycle, similarly as in \cite{noh}, where authors considered a simpler case $f_i=1$. 
	
	The Shannon entropy (per particle) of the equilibrium distribution $S_\infty=-\sum_{k=1}^M\mu_k\log\mu_k$  can be rewritten  as 
	\begin{equation}\label{eq:S1}
	S_\infty=-\frac{(f_ig_i\log(f_ig_i))_{\mathrm{av}}}{(f_ig_i)_{\mathrm{av}}}+\log(f_ig_i)_{\mathrm{av}}+\log M,
	\end{equation}
	where $( \cdot )_{\mathrm{av}}$ denotes average density over all nodes in the graph. The term $\log M$ in  (\ref{eq:S1}) corresponds to the maximal entropy of the system when  particles  are uniformly distributed, i.e.  $\bar{\mu}_i=1/M$ .   It is easy to prove that $S_\infty$ is always less than $\log M$, which is a simple consequence of the convexity of function $f(x)=x\log x$ and Jensen's inequality \cite{cover}. The non-positive valued sum of the first two terms on the rhs (\ref{eq:S1}) 
	describes the lowering of equilibrium particle entropy induced by the complexity of network topology. It equals to zero when every node possesses the same fitness factor $f_i=f$ and the network  is a regular graph, i.e., $k_i=k_{av}$.
	
\begin{figure}[t]
		\flushright{		 \epsfig{file=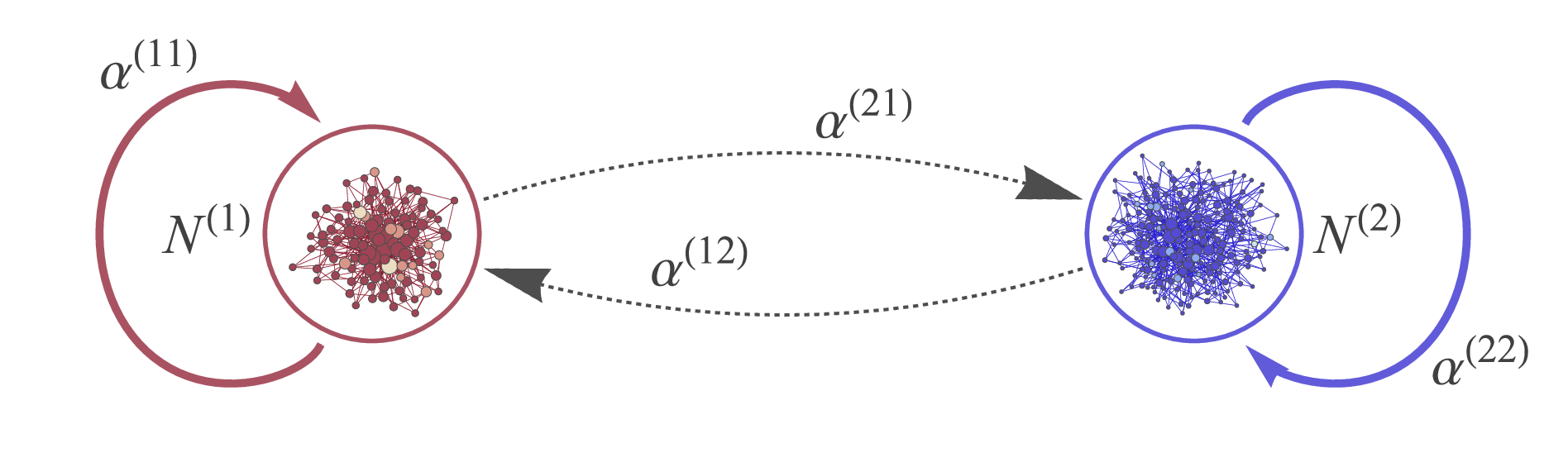,width=0.5\columnwidth}}
		\flushleft{\vspace{-3.15cm}\epsfig{file=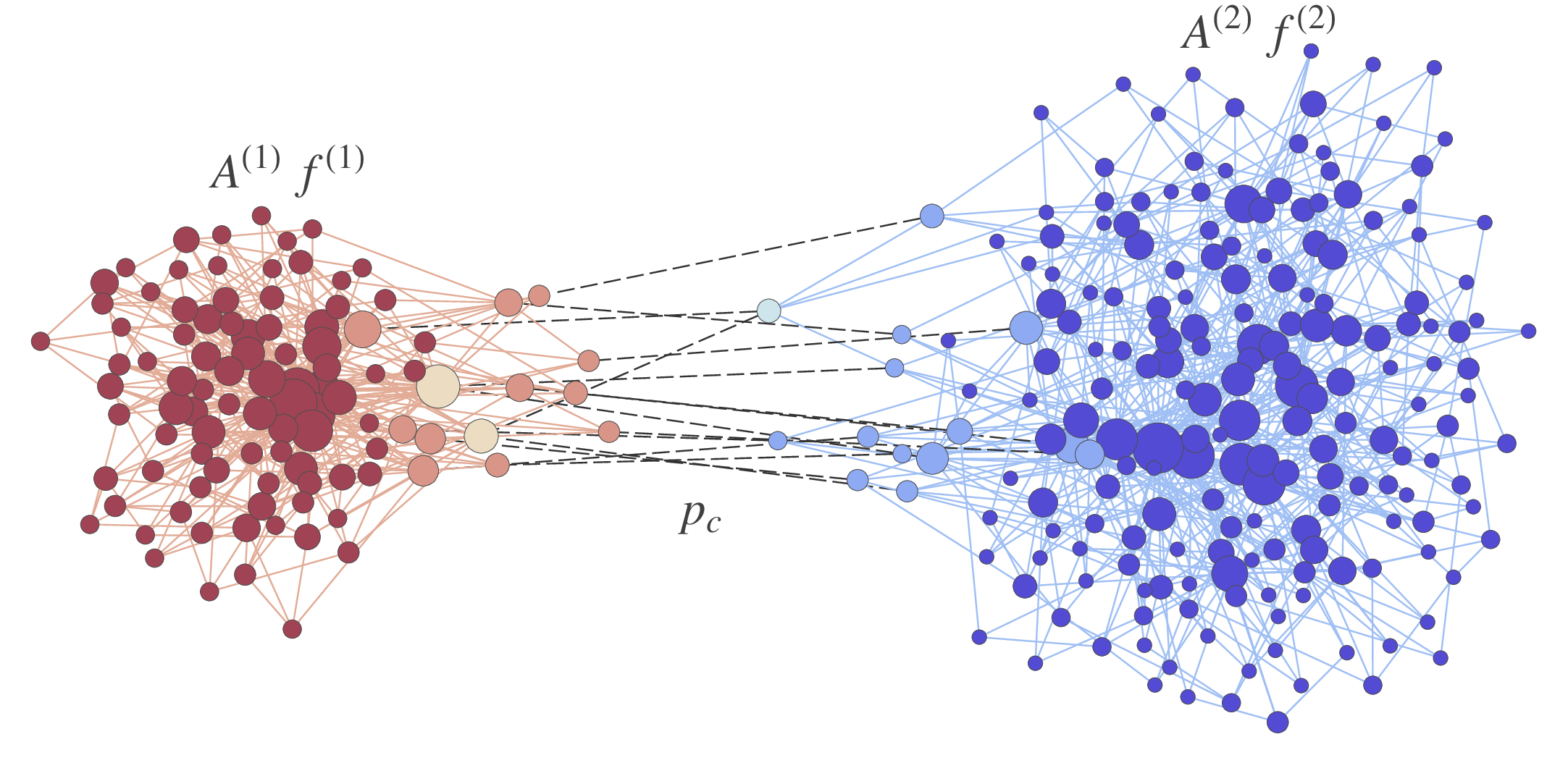,width=0.5\columnwidth}}
		
		\caption{Weakly coupled complex networks. Fig. (a) presents two connected Barab\'asi-Albert sub-networks with sparse inter-links controlled by the parameter $p_c$. Fig. (b) illustrates the approximation of the time-scale separation, which is equivalent to substituting the pair of connected sub-networks by a graph consisting of two nodes.}
		\label{fig:fig1}
	\end{figure}

	Let two weakly connected networks $\mathcal{A}^{(1)}$ and $\mathcal{A}^{(2)}$ (see Fig. \ref{fig:fig1} (a)) sized $M^{(1)}$ and $M^{(2)}$ respectively, be represented as sub-networks  of  $\mathcal{A}$. Internal connections between nodes within the same sub-network $\mathcal{A}^{(a)}$, $a=1,\,2$ are defined by adjacency matrices $A^{(a)}\in\mathbb{M}^{M^{(a)}\times M^{(a)}}(\{0,\,1\})$, while the matrix $C\in\mathbb{M}^{M^{(1)}\times M^{(2)}}$ represents links between these two sub-networks. The global adjacency matrix is
	\begin{equation}\label{eq:matrix}
	A_{\mathrm{total}}=\left[\begin{array}{cc}A^{(1)}&C\\C^T&A^{(2)}\end{array}\right].
	\end{equation} 
	
	We assume that links between the sub- networks $\mathcal{A}^{(1)}$  and $\mathcal{A}^{(2)}$ are defined by a  matrix   $C$ that is a sparse random matrix described by the  parameter $p_c \ll 1$ equal to  the probability that two nodes belonging to different sub-networks $\mathcal{A}^{(1)}$ and $\mathcal{A}^{(2)}$ are directly connected. The global vector of attractiveness $\mathbf{f}\in \mathbb{R}_{+}^{M^{(1)}+M^{(2)}}$ consists of corresponding vector components $\mathbf{f}^{(a)}\in \mathbb{R}_{+}^{M^{(a)}}$ that signify the attractiveness of nodes in separate sub-networks. 
	
	\section{Analytical results for coarse-grained diffusion} \label{SecIII}
	
	We will assume that networks $\mathcal{A}^{(1)}$  and $\mathcal{A}^{(2)}$ are {\it weakly coupled} i.e. the parameter $p_c$ is so small that  the flow of particles between the node $i$ in the sub-network $\mathcal{A}^{(a)}$ and all other nodes within this sub-network is much greater than the flow from this node to another sub-network $\mathcal{A}^{(b)}$. This  happens when the attractiveness $g_i^{(a)}$ of a node neighborhood in the sub-network $a$, before it is connected, is far greater than the attractiveness of node neighborhood in the sub-network $b$ 
	\begin{align}
	g^{(a)}_i\gg p_cM^{(b)} f^{(b)}_{\mathrm{av}},\,\, i=1,\,2,\,\dots,\,M^{(a)}.\label{eq:cond_1}
	\end{align}
	When  flows within network   $\mathcal{A}^{(1)}$ and  within  network  $\mathcal{A}^{(2)}$ are greater than flows between these networks, then the characteristic time scales for internal processes are shorter  than for processes that take place between networks. Subsequently, one can apply the  adiabatic approximation \cite{adiab} for the number of particles in the $i$-th node of the $a$-th network at a time moment $t$  as $n^{(a)}_i(t)=N^{(a)}(t)\mu^{(a)}_i$ $i=1,\,\dots,\,M^{(a)}$. Here $N^{(a)}(t)$ is the total number of particles in the $a$-th network at time $t$ and $\mu^{(a)}_i$ is the equilibrium distribution of density of particles in the non-connected networks given by Eq. (\ref{eq:n_i}).
	The separation of time scales, which follows from the assumption in (\ref{eq:cond_1}), is related to Cheeger's inequality for Markov Chains as discussed in \cite{lawler}. In fact, the second eigenvalue of the Markov operator (describing time scale of the diffusion process) is bounded by a constant that is determined by the slowest transition between one sub-network and its complement in the graph. 
	Suppose that the fitness $f_i$ of each node is  equal to $f>0$. Then the average attractiveness of a node neighborhood in a network $a$ is equal to $(g^{(a)})_{\mathrm{av}}=f k^{(a)}_{\mathrm{av}} $, $a=1,\,2$. The relations (\ref{eq:cond_1}) lead to the necessary condition for our approach 
	\begin{equation}\label{eq:comp_degrees}
	k^{(a)}_{\mathrm{av}} \gg M^{(b)} p_c,
	\end{equation}
	where $a,\,b=1,\,2$, $a\neq b$. It means that the average {\it inter-network degree} must be smaller than the average {\it internal degree}. It is worthwhile to check   validity of the above mentioned condition  when the connected networks are two Erd\H{o}s-R\'{e}nyi (ER) graphs \cite{ER},  two Barab\'asi-Albert (BA) networks \cite{Barabasi,Barabasi2} or two Watts-Strogatz graphs (WS) \cite{Watts}. Let us assume that in the first case  the graphs possess equal numbers of nodes ($M^{(1)}=M^{(2)}$) and that these graphs are described by the parameters $p^{(1)}$ and $p^{(2)}$,  which correspond to the probability that any two randomly chosen nodes will be directly connected in a given graph. We then get
	\begin{equation}\label{eq:condER}
	p_c \ll p^{(1)}, p^{(2)},
	\end{equation}
	and it is clear that these conditions are independent of the network size, which means we can use the same $p_c$ for networks of all sizes. The approach is justified because the dispersion of degree distribution is negligible. In the case of BA networks  of sizes $ M^{(1)}=M^{(2)}=M$ and  described by  parameters $m^{(1)}$ and $m^{(2)}$ corresponding to  numbers of links every new node is attached in the evolutionary BA model \cite{Barabasi,Barabasi2}, the situation is different since the degree distribution is a power law. An approximate condition corresponding to (\ref{eq:condER})  is related to the size $M$ of  network and   can be obtained by taking the  average node degree in a BA network  $k^{(a)}_{\mathrm{av}BA}=2m^{(a)}$ thus  
	\begin{equation}\label{eq:condBA}
	p_c \ll \frac{2m^{(1)}}{M}, \frac{2m^{(2)}}{M}.
	\end{equation}
	For two connected WS graphs with equal numbers of nodes  $M^{(1)}=M^{(2)}=M$, average degrees  $ K^{(1)}=K^{(2)}=K$ and parameters of randomness (see \cite{Watts}) $\beta^{(1)},\,\beta^{(2)}$  Eq. (\ref{eq:comp_degrees})  takes the following form
	\begin{equation}
	p_c \ll \frac{K}{M},
	\end{equation}
	and it does not depend on the values of parameters $\beta^{(1)},\,\beta^{(2)}$. This is a completely different situation than in the recent work of Jedrzejewski \cite{Jedrzejewski}, in which the author  presented that his approach for the q-voter model \cite{Javarone,Chmiel,Mellor} fails for the small values of parameters $\beta$. In our approach  more important than an internal structure of the network is its average degree, which, in the case of WS graphs, is equal $K$, and does note depend on other parameters. The above means that the our method  works uniformly for every value of $\beta$, as long as the ratio $K/M$ is sufficiently small.

	Using the above approximation we get from Eq. (\ref{eq:dynamics})
	\begin{equation}\label{eq:NNON}
	N^{(a)}(t+1)=\sum_{b=1}^{2}\alpha^{(a\,b)}N^{(b)}(t),
	\end{equation}
	where
	\begin{equation}\label{eq:able2}
	\alpha^{(a\,b)}=p_c^{(a\,b)}M^{(a)}\frac{f^{(a)}_{\mathrm{av}}f^{(b)}_{\mathrm{av}}}{(f^{(b)}_ig^{(b)}_i)_{\mathrm{av}}},\;\alpha^{(a\,a)}=1-\sum_{r\neq a}\alpha^{(r\,a)}.
	\end{equation}
	where  $p_c^{a,b}=p_c$.
	Parameters $\alpha^{(1\,2)}$ and $\alpha^{(2\,1)}$ describe integrated transition probabilities between the networks at the macroscopic level (see Fig. \ref{fig:fig1}). If there is only a negligible correlation between node degree and node fitness factor  i.e. $(f^{(a)}_ig^{(a)}_i)_{\mathrm{av}} \approx (f^{(a)}_{\mathrm{av}})^2k^{(a)}_{\mathrm{av}}$   then 
	\begin{equation} \label{alpha}
	\alpha^{(a\,b)}\approx p_{c} M^{(a)}f^{(a)}_{\mathrm{av}}/(f^{(b)}_{\mathrm{av}}k^{(b)}_{\mathrm{av}}).
	\end{equation}

	It follows that the equilibrium number of particles in the network $\mathcal{A}^{(a)}$ is : $N^{(a)}_\infty=N \alpha^{(a\,b)}/(\alpha^{(a\,b)}+\alpha^{(b\,a)})$. Near the  equilibrium, the variations of $N^{(a)}(t)$ are sufficiently slow thus allow to write Eq. (\ref{eq:NNON}) in the form that is analogous to Ficks's First Law \cite{mazo} with a diffusion constant $D$ and in the presence of an additional inter-network driving force $F$
	\begin{equation}\label{eq:fick}
	\dot{N}^{(1)}(t)=-\left(N^{(1)}(t)-N^{(2)}(t)\right)D+ F N,
	\end{equation}
	where 
	\begin{equation}\label{eq:10x}
	D=(\alpha^{(1\,2)}+\alpha^{(2\,1)})/2\;\;\;\mathrm{and}\;\;\; F=(\alpha^{(1\,2)}-\alpha^{(2\,1)})/2.
	\end{equation}
	
	It is important to note that driving force $F$ can cause a non-zero flow $\dot{N}^{(1)}(t)$ even if there is no gradient $N^{(1)}(t)-N^{(2)}(t)$. In fact the force $F$ results from two reasons: differences in network sizes $M^{(a)}$; differences in network mean fitnesses ${f}^{(a)}_\mathrm{av}$ and differences in mean network degrees $k^{(a)}_{\mathrm{av}}$. The last dependency means that the internal network structure influences the inter-network diffusion, thus a sub-network can be attributed a potential related to its topology.
	
	One can see  from Eq. \ref{alpha} that for a given sub-network size  the denser is the sub-network,  the greater the number of particles $N^{(a)}_{\infty}$ in this sub-network  in the equilibrium state.  in fact adding  new links between nodes inside  a  given sub-network $\mathcal{A}^{(1)}$ increases its mean degree $k^{(1)}_{\mathrm{av}}$, thus parameter $\alpha^{(2\,1)}$ decreases and more particles are retained in the network.
	It follows that alternations  of internal topologies of sub-networks can induce or  reverse a particle flow between them.
	It is interesting from a practical point of view that by changing structure of  sub-networks  one can not only modify diffusion constants but also induce or reverse  the direction of the particles' flow (see Sec. \ref{Sec_postrec}).
	\section{Numerical simulations of coarse-grained diffusion}  \label{SecIV}
	\begin{figure}[t!]
		\centerline{\epsfig{file=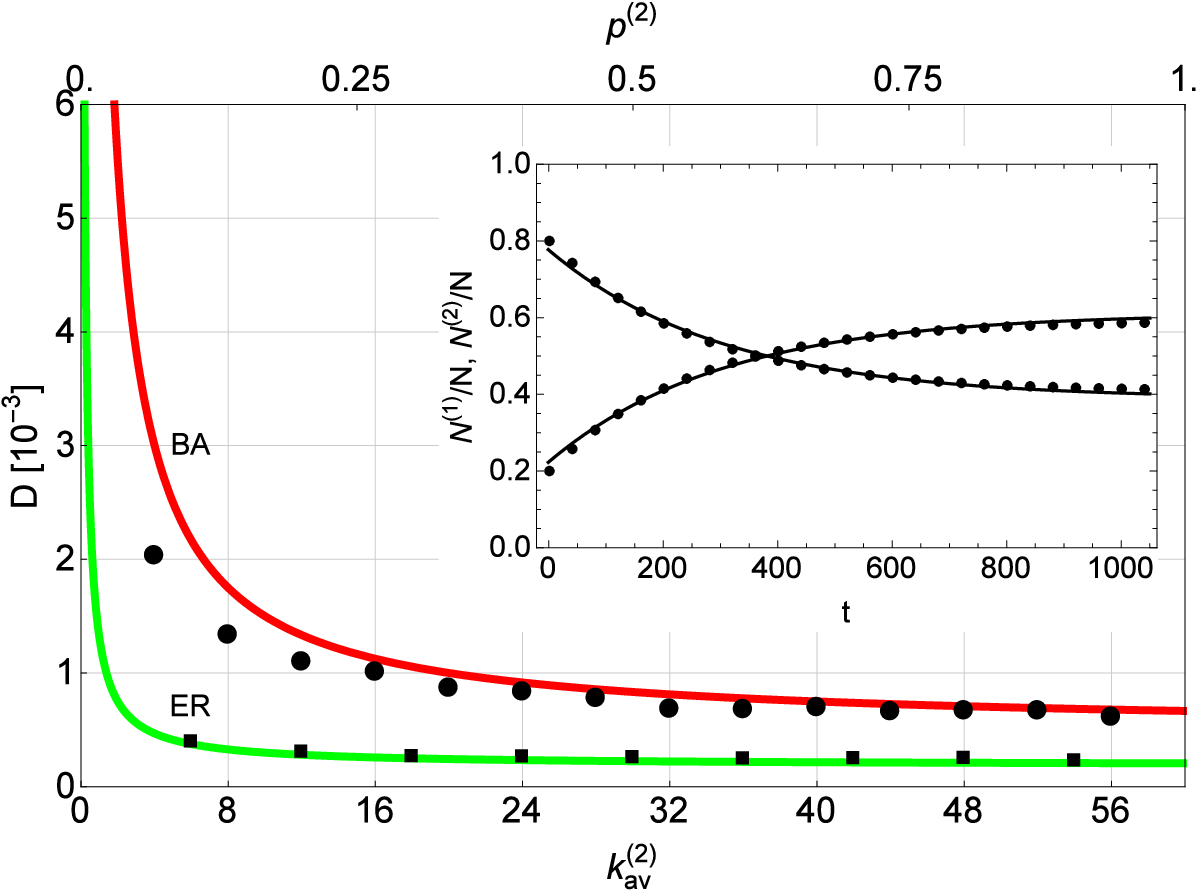,width=0.6\columnwidth}}
		
		\caption{Influence of internal link density of a given sub-network on a diffusion constant between two weakly-connected complex networks. The red line (analytical solution) and the small solid circles (numerical simulations) correspond to a diffusion constant of a pair of BA networks as function of $k^{(2)}_{\mathrm{av}}$ (see the bottom frame). The green line (analytical solution) and the squares (numerical simulations) correspond to the diffusion constant of a pair of ER sub-graphs as  function of $p^{(2)}$ (see the upper frame).  The inset shows a comparison of  the solution (\ref{eq:NNON}) and  a numerical simulation for  evolution of numbers of particles $N^{(1)}(t)$, $N^{(2)}(t)$, in two BA sub-networks. $500$  copies of corresponding networks have been used to reach appropriate mean values of diffusion constants $D$.}
		\label{fig:fig2}
	\end{figure}
	\begin{figure}[b!]
		\centerline{\epsfig{file=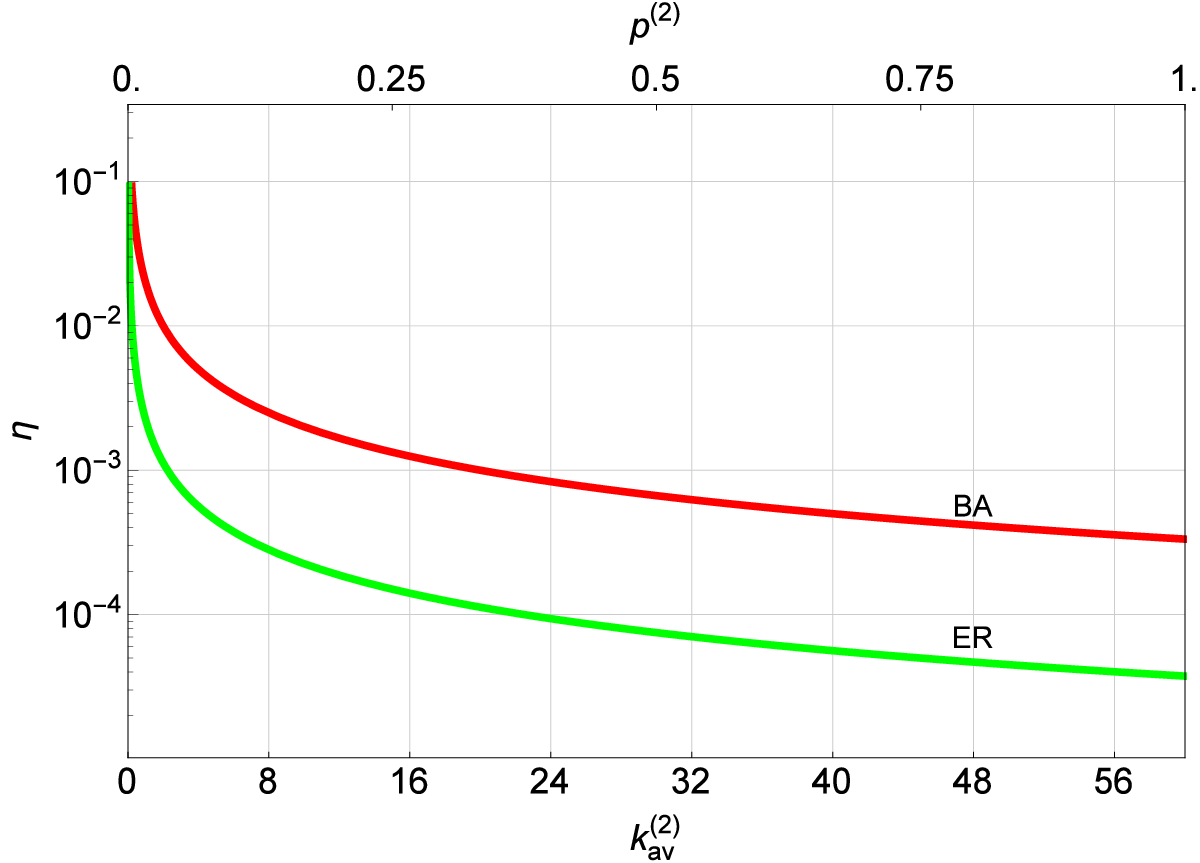,width=0.6\columnwidth}}
		
		\caption{Validity of approximations (\ref{eq:condBA}) -- red curve (BA networks, see the bottom frame) and  (\ref{eq:condER}) -- green curve (ER graphs, see the top frame). The variable $\eta$ displayed at the vertical axis is the ratio between the mean node degree for inter-network connections between two sub-networks and the mean node degree for internal connections in a given  sub-network. Its value is $\eta_{\mathrm{BA}}=p_cM/k_{\mathrm{av}}^{(2)}$ for BA networks and  $\eta_{\mathrm{ER}}=p_c/p^{(2)}$ for  ER graphs. The condition  $\eta\ll 1$ is equivalent to Eqs. (\ref{eq:condER}) and (\ref{eq:condBA}). Network's parameters are the same as for Fig. \ref{fig:fig2}.}
		\label{fig:fig3}
	\end{figure}
	All  numerical simulations were implemented in the Wolfram language with Wolfram Mathematica environment of the version  11.2 with the   standard \texttt{MachinePrecision} ($16$ digits). With such a setting a typical personal computer has enough computing power for handling the scripts.  Fig. \ref{fig:fig2} presents the influence of internal link density of a given sub-network on a diffusion constant for two connected Barab\'asi-Albert networks ($M^{(1)}=M^{(2)}=500$, $k_\mathrm{av}^{(1)}=20$, $p_c=4\cdot 10^{-5}$) and the two connected Erd\H{o}s--R\'{e}nyi graphs ($M^{(1)}=M^{(2)}=200$, $p^{(1)}=0.1$, $p_c=3.75\cdot 10^{-5}$). For each case $f^{(1)}_{i_1}=f^{(2)}_{i_2}={const}$. Analytical calculations following from Eqs. (\ref{eq:able2}) and (\ref{eq:fick}) correspond well with the numerical results obtained from random-walk simulations. Parameters $\alpha^{(a\,b)}$ and the resulting diffusion constant $D$ were obtained from the best-fit solution of Eq. (\ref{eq:NNON}) to  $N^{(a)}(t)$ - see inset in Fig \ref{fig:fig2}. Fig. \ref{fig:fig2} confirms  that the diffusion constant $D$ decreases as the internal density of a network increases. It is understandable that when a network becomes denser, it retains particles in its interior more easily since a particle is more likely to choose internal rather than external links, thus as result the inter-network diffusion coefficient $D$ decays.

	Differences between numerical simulations and analytical results observed at  Fig. \ref{fig:fig2} for the BA networks  in the regime of small $k_{\mathrm{av}}^{(2)}$ may be understood looking at the  Fig. \ref{fig:fig3}. It  presents the  ratio $\eta= p_cM^{(b)} f^{(b)}_{\mathrm{av}}/g^{(a)}_i$ between the mean node degree for connections to another network and the mean node degree for internal connections   as functions  of the same networks parameters as at Fig. \ref{fig:fig2}. In fact the condition $\eta\ll 1$ is equivalent to  the weakly coupling condition (\ref{eq:cond_1}) that   simplifies to Eq. (\ref{eq:condBA}) for BA networks and to (\ref{eq:condER}) for ER graphs. Numerical and analytical values of diffusion  constant for  ER graphs presented at Fig.~\ref{fig:fig2} fit very well  in the whole range of the parameter  $p^{(2)}$ and the corresponding parameter $\eta_{ER}$ depicted at Fig. \ref{fig:fig3} is smaller than $3\cdot 10^{-4}$ in this region. On the other hand  numerical and analytical values of the diffusion  constant for BA networks  presented at Fig.  \ref{fig:fig2} show some disagreement when the internal degree for BA networks $k_\mathrm{av}^{(2)} = 4$.     However the corresponding parameter $\eta_{BA}$ depicted at Fig. \ref{fig:fig3} for $k_\mathrm{av}^{(2)}=4$  is about  $0.5 \cdot 10^{-2}$ so it is much larger than for ER graphs. 
	

	\begin{figure}[ht!]
		\centerline{\epsfig{file=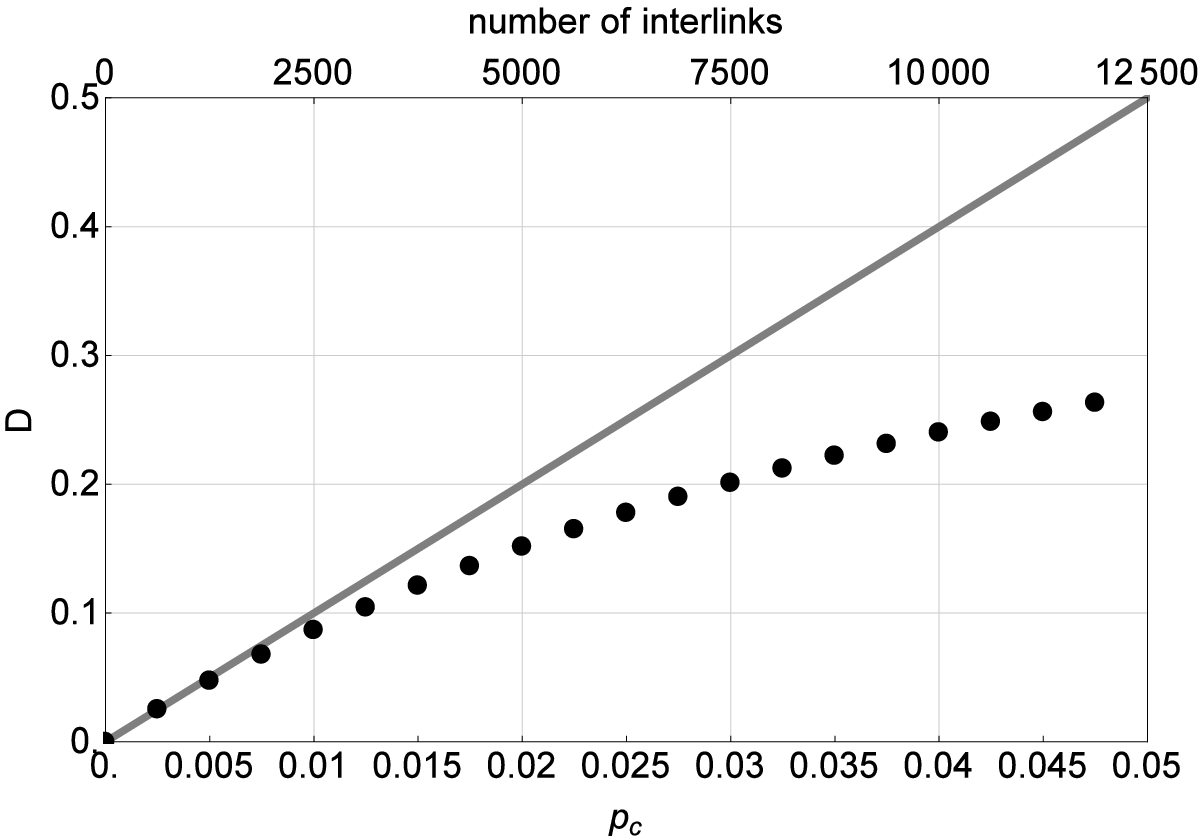,width=0.6\columnwidth}}
		
		\caption{ Dependence of diffusion coefficient $D$  on the coupling parameter $p_c$ for two BA networks with parameters $M=M^{(1)}=M^{(2)}=500$, $f^{(1)}_{i_1}=f^{(2)}_{i_2}=const.$ and $m=m^{(1)}=m^{(2)}=25$. Circles correspond to simulation results and the solid line to Eq. (\ref{eq:able2}). The fitting fails around $p_c \approx 0.01$  while the approximate condition  (\ref{eq:condBA}) for the validity of our analytical approach for this system can be written as   $p_c \ll  2m/M=0.1 $. The upper abscissa shows the number of interlinks given by the formula $p_c M^2$. 
			Let us note since the  density  internal links in every sub-network is high  (the average  degree equals to $10$ ) thus  the analytical results well fit to numerical simulations even if the number of interlinks is larger than $1000$, i.e when every node in one sub-network possesses in average two interlinks to another sub-network.    }
		
		\label{fig:fig6}
	\end{figure}

	
	Fig. \ref{fig:fig6} presents a complementary study of the diffusion constant $D$  for coupled BA networks when the density of interlinks $p_c$ is changed. One can see that the analytical approach works properly when the parameter $p_c$ is at least one order of magnitude smaller than the critical parameter value estimated  from  Eq. (\ref{eq:condBA}).

	\begin{figure}[ht!]
		\centerline{\epsfig{file=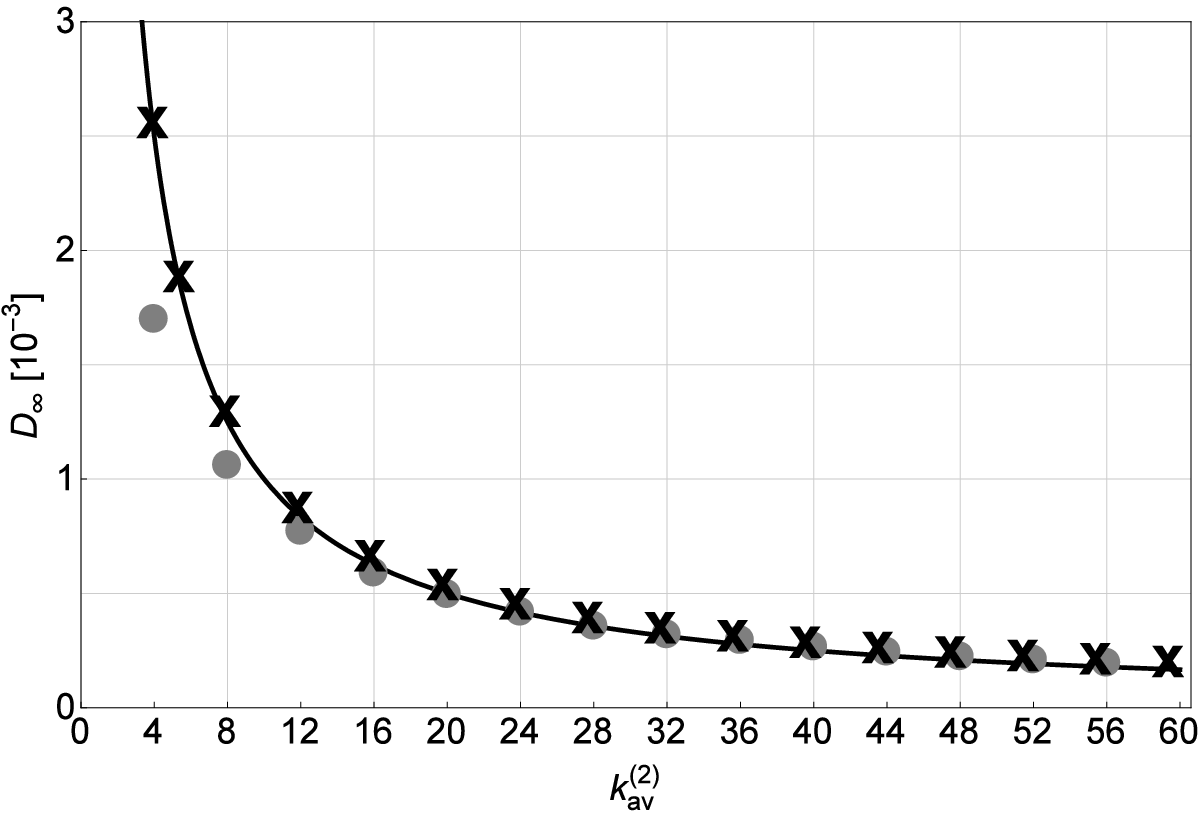,width=0.6\columnwidth}}
		
		\caption{Asymptotic diffusion constant for two BA networks ($M^{(1)}=M^{(2)}=500$, $f^{(1)}_{i_1}=f^{(2)}_{i_2}={const.}$ and $p_c=4 \cdot 10^{-5}$) in the case when the first of them is much  denser than the second one $k_\mathrm{av}^{(1)}=200 \gg k_\mathrm{av}^{(2)}$.   Grey circles are results of numerical simulations, the solid black line comes from Eqs. (\ref{eq:able2}) and (\ref{eq:10x}) and crosses correspond to the approximation  (\ref{eq:approx}) when the diffusion constant is independent from the mean node's degree $k_\mathrm{av}^{(1)}$ in the first network. One can see that this approximation is very good for     $k_\mathrm{av}^{(2)}  > 8$ and the diffusion constant in this region  is inversely proportional to the mean node degree in the second network. When the second network is too sparse   the condition  (\ref{eq:condBA}) required  for the time scales separation is not fulfilled.}
		\label{fig:fig5}
	\end{figure}
	
	When both sub-networks are of comparable  sizes but  the network $\mathcal{A}^{(1)}$   is much  denser  then  it  keeps majority of walkers in its interior and prevents the diffusion between networks.
	For fitness parameters $f^{(1)}_{i_1}=f^{(2)}_{i_2}=f$ one gets the asymptotic diffusion constant as
	\begin{equation}\label{eq:approx}
	D_\infty=\frac{p_c}{2}\left(\frac{M^{(1)}}{k_\mathrm{av}^{(2)}}+\frac{M^{(2)}}{k_\mathrm{av}^{(1)}} \right)\approx \frac{p_c}{2}\frac{M^{(1)}}{k_\mathrm{av}^{(2)}}.
	\end{equation}
	Fig. \ref{fig:fig5} demonstrates that   the result (\ref{eq:approx}) fits well to  numerical simulations for coupled   BA networks.
	
	\section{Predicting direction of  diffusion flux } \label{Sec_postrec}
	Now we apply our formalism  to  predict   a  flux direction of  diffusive   particles that are initially evenly spread  over   two weakly connected networks.   In other words  we find which of the networks is more attractive.     We consider a class of models  where    nodes'  fitness  factors are  powers of their internal degrees i.e.
	\begin{equation}\label{eq:fik}
	f_i^{(a)}=\left(k^{(a)}_i\right)^\beta,
	\end{equation}
	The characteristic exponent $\beta$ is a model  parameter. When  $\beta>0$  then nodes with larger degrees are more  attractive for diffusing particles  than nodes with smaller degrees that can be considered as a kind of preferential diffusion.    The condition   $\beta<0$  corresponds to the opposite situation (anti-preferential diffusion)  and the case  $\beta=0$  is equivalent to  the standard random walk without preferred nodes. 
	
	When $N^{(1)}= N^{(2)}$ then the diffusion direction follows from the sign of the driving force $F$ (see Eq. \ref{eq:10x}). Using  Eq. (\ref{eq:fik}) we get
	\begin{equation} \label{eq:force}
	F=\frac{\alpha^{(12)}-\alpha^{(21)}}{2}=\frac{p_c}{2}
	\left[
	\frac{M^{(1)}\left(
		(k^{(1)})^\beta\right)_\mathrm{av}}     { \left(
		(k^{(2)})^{\beta+1}\right)_\mathrm{av}}  
	-   \frac{M^{(2)}\left(
		(k^{(2)})^\beta\right)_\mathrm{av}}     { \left(
		(k^{(1)})^{\beta+1}\right)_\mathrm{av}}  
	\right].       
	\end{equation}
	Condition for the equilibrium is equivalent to $F=0$, i.e.
	
	\begin{equation}
	M^{(1)}\left((k^{(1)})^\beta\right)_\mathrm{av}\left((k^{(1)})^{\beta+1}\right)_\mathrm{av}\label{eq:condition} =M^{(2)}\left((k^{(2)})^\beta\right)_\mathrm{av}\left((k^{(2)})^{\beta+1}\right)_\mathrm{av}.
	\end{equation}
	For  networks of  the same sizes diffusing particles will move  to a network with a higher product of moments  $\beta$ and $\beta+1$ of their nodes degrees. For $\beta=0$ it means  a flux towards  a denser network. On the other hand for two networks with the same degree distributions the larger one is more attractive. For  a  better illustration of   the meaning of  condition (\ref{eq:condition}) we consider a few special cases.
	
	\subsection{Preferential diffusion $(\beta=1)$ at two networks with the same number of nodes and average degree  }
	When   both networks  possess the same numbers of nodes $M^{(1)}=M^{(2)}=M$ , the same average degrees  $\left(k^{(1)}\right)_\mathrm{av}=\left(k^{(2)}\right)_\mathrm{av}=k_\mathrm{av}$  and  $\beta=1$ we get  
	
	\begin{equation}F=\frac{p_cMk_\mathrm{av}}{2}\left[\frac{1}{\left((k^{(2)})^2\right)_\mathrm{av}}-\frac{1}{\left((k^{(1)})^2\right)_\mathrm{av}}\right].\end{equation}
	In such a case  the  network with a higher variance of degree distributions is more attractive, i.e.  it collects a larger number of  particles 
	
	\subsection{Anti-preferential diffusion  $(\beta=-1)$}
	
\begin{figure}[ht!]
		\centerline{\epsfig{file=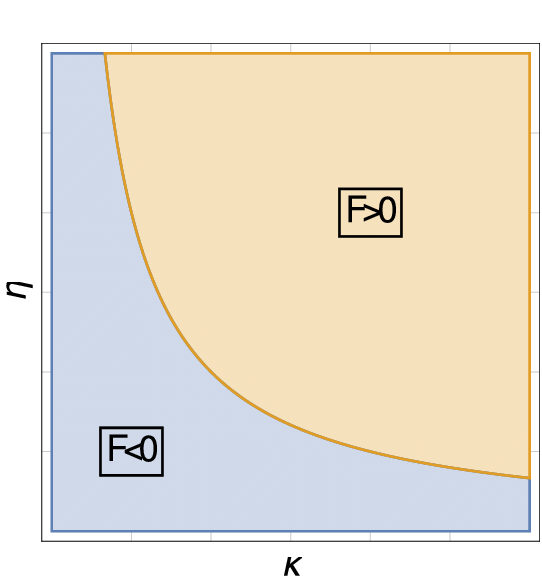,width=0.35\columnwidth}\;\;\;\;\;\;\;\;\;\;\;\;\;\;\;\;	\epsfig{file=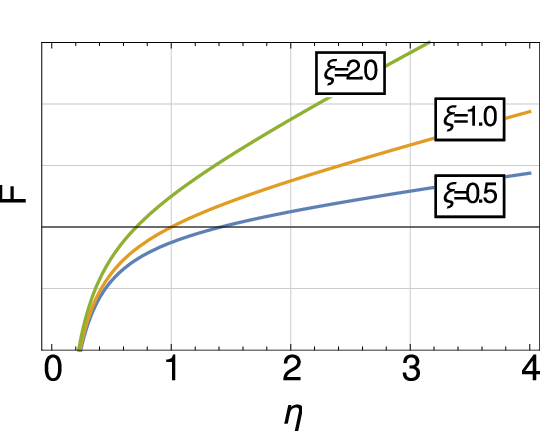,width=0.43\columnwidth}}
		
		\caption{Left panel: signs of driving force for networks  for  anti-preferential diffusion  $(\beta=-1)$ as functions of model parameter, see Eq. (\ref{eq:etakappa}) . Right panel: driving force for two ER graphs with standard diffusion  ($\beta=0$), see Eq. (\ref{eq:xieta}).}
		\label{fig:figR1}
	\end{figure}

	When the exponent $\beta=-1$ then   the condition for the  flux balance  (\ref{eq:condition}) simplifies to
	\begin{equation}
	M^{(1)}\left((k^{(1)})^{-1}\right)_\mathrm{av}=M^{(2)}\left((k^{(2)})^{-1}\right)_\mathrm{av}\;\;
	\Rightarrow\;\;{M^{(1)}}{M^{(2)}}=\frac{\left((k^{(2)})^{-1}\right)_\mathrm{av}}{\left((k^{(1)})^{-1}\right)_\mathrm{av}}\;\;
	\Rightarrow\;\;
	\eta=\kappa^{-1},\label{eq:etakappa}
	\end{equation}
	where  $\eta=M^{(1)}/M^{(2)}$ and $\kappa=\left((k^{(1)})^{-1}\right)_\mathrm{av}/\left((k^{(2)})^{-1}\right)_\mathrm{av}$ . The result is presented at  Fig.  \ref{fig:figR1}) . 
	
	\subsection{Standard diffusion $(\beta=0)$ at two ER graphs }
	In the case of  standard diffusion    $\beta=0$    values of  fitness function  $f_i$ of any node are  the same. For two  connected ER graphs with   average degrees y  $k^{(a)}_\mathrm{av}=p^{(a)}M^{(a)}$    the driving force is  
	\begin{equation}
	F=\frac{p_c}{2}\left[\frac{M^{(1)}}{p^{(2)}M^{(2)}}-\frac{M^{(2)}}{p^{(1)}M^{(1)}} \right]=\frac{p_c}{2p^{(1)}}\left[\frac{p^{(1)}M^{(1)}}{p^{(2)}M^{(2)}}-\frac{M^{(2)}}{M^{(1)}} \right],
	\end{equation}
	Introducing the parameter  $\xi=p^{(1)}/p^{(2)}$ we get 
	\begin{equation}\label{eq:xieta}
	F\propto \left(\xi\eta-\eta^{-1}\right).
	\end{equation}
	
	The relation is illustrated at Fig. \ref{fig:figR1}.

	\subsection{Strongly preferential diffusion $(\beta > 2)$ at two BA networks}
	For a BA network of size $M^{(a)}$ and a  mean degree $2m^{(a)}$  values of moments $\left((k^{(a)})^\beta\right)_{\mathrm{av}}$  (for   $\beta\neq 2$)  are equal to  
	\begin{equation}
	\left((k^{(a)})^\beta\right)_{\mathrm{av}}\approx \frac{2\left(m^{(a)}\right)^\beta}{2-\beta}\left[\left(\frac{m^{(a)}}{M^{(a)}}\right)^{2-\beta}-1\right],
	\end{equation}
	Assuming that  both networks  are large i.e. $M^{(1)},M^{(2)}\gg 1$) and $\beta <2$, $\beta\neq 1$  we get the following condition  for vanishing of driving force   \ref{eq:force}
	
	\begin{equation}
	\left(m^{(1)}\right)^{2\beta+1}\left[\left(\frac{m^{(1)}}{M^{(1)}}\right)^{3-2\beta}+1\right]= \left(m^{(2)}\right)^{2\beta+1}\left[\left(\frac{m^{(2)}}{M^{(2)}}\right)^{3-2\beta}+1\right].
	\end{equation}

	\section{Entropy production in coupled networks} \label{SecV}

	Using the above discussed  adiabatic approximation the total system entropy can be written after some algebra  as
	\begin{align}
	S_{\mathrm{total}}(t)=\left[N^{(1)}(t)S_\infty^{(1)}+N^{(2)}(t)S_\infty^{(2)}\right]+\label{eq:entro}
	\left[-N^{(1)}(t)\log N^{(1)}(t)-N^{(2)}(t)\log N^{(2)}(t)+N\log N\right],
	\end{align}
	where $S_\infty^{(a)}$ is the equilibrium entropy per particle for a  network $\mathcal{A}^{(a)}$ as given by (\ref{eq:S1}). 
	This result has a  clear physical implications. The term in the first square bracket in (\ref{eq:entro}) follows from the lack of {\it microscopic} information about a particle's position {\it inside} every  network and can be viewed as a  network's internal entropy. The term in the second bracket is a {\it macroscopic} entropy of the two-state Markov system. It results  because of the missing information about the particles distribution {\it between} both sub-networks. Time changes of  the system entropy  can be written as
	
	\begin{equation}
	\sigma(t)=\dot{N}^{(1)}(t)\left(S_\infty^{(1)}-S_\infty^{(2)}+\log N^{(2)}(t)-\log N^{(1)}(t)\right) 
	\end{equation}
	and they follow from the transport of microscopic entropies related to sub-network topologies $S_\infty^{(a)}$ (that can vary in both networks) and from the entropy changes at the macroscopic level. 
	
	\section{Case of many coupled networks} \label{SecVI}
	
	Until now we have considered  models  of two coupled sub-networks. We can easily demonstrate (see. \ref{sec:non}) that our approach is valid for any system of $m$ weakly coupled networks. Then, Eq. (\ref{eq:NNON}) includes a sum over $m$ networks and $p_c^{(a\,b)}$ is a symmetrical inter-network connectivity matrix (with zeros at the diagonal and small off-diagonal elements). Assuming that our system of weakly coupled networks  forms a connected graph one gets  instead of  Eq. (\ref{eq:fick}) the following coarse-grained equation for number of particles in each network  
	\begin{align}
	\dot{N}^{(a)}(t)=\sum_{b\neq a}[-&D^{(b\,a)}(N^{(a)}(t)-N^{(b)}(t))+ F^{(b\,a)}(N^{(a)}(t)+N^{(b)}(t)) ],
	\end{align}
	where 
	\begin{equation}
	D^{(a\,b)}=\frac{1}{2}(\alpha^{(a\,b)}+\alpha^{(b\,a)})\;\;\; \mathrm{and\;\;\;} F^{(a\,b)}=\frac{1}{2}(\alpha^{(b\,a)}-\alpha^{(a\,b)}).
	\end{equation}
	The corresponding entropy production can be written in a compact form as 
	\begin{equation}
	\sigma(t)=\sum_{a=1}^{m}\dot{N}^{(a)}[S^{(a)}_\infty-\log N^{(a)}(t)].\label{eq:entro3}
	\end{equation}
	
	\section{Diffusion on weighted and hierarchical networks} \label{SecVII}
	
\begin{figure}[ht!]
		\centerline{\epsfig{file=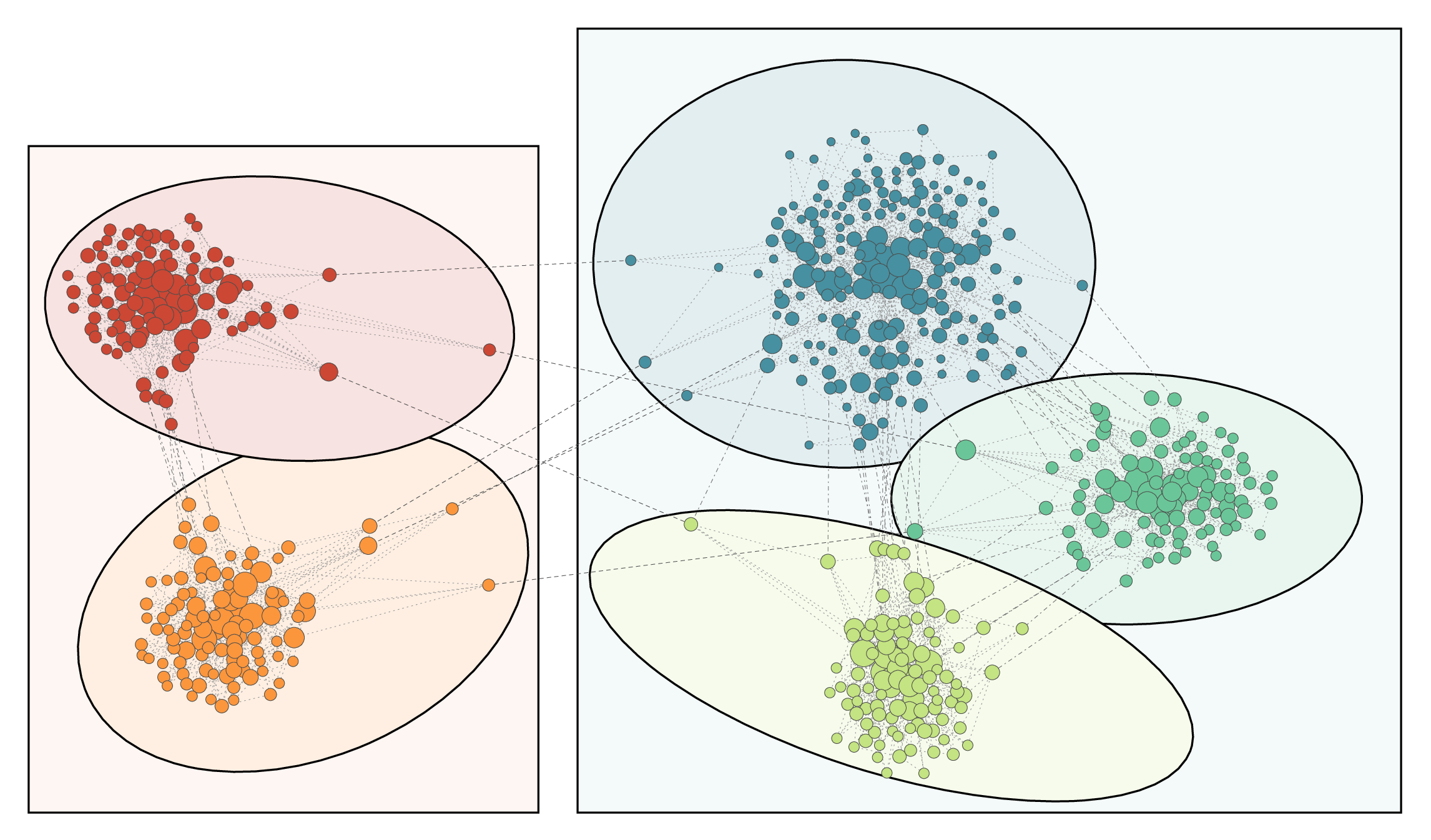,width=0.8\columnwidth}}
		
		\caption{Hierarchical system of weakly coupled networks with  three levels of nested hierarchies.}
		\label{fig:fig4}
	\end{figure}

	The approach developed in Sec. \ref{SecIII} is also valid if one considers diffusion on {\it weighted networks} with self-loops. In such a  case, the symmetrical adjacency matrix $A$ in Eq. (\ref{eq:dynamics}) should be replaced by a corresponding adjacency matrix with non-negative elements including the diagonal ones. This observation allows us to extend this framework  further to {\it nested hierarchical systems} (see e.g. \cite{hierach,Ravasz,Sales-Pardo,Clauset} and Fig.~\ref{fig:fig4}).
	
	To illustrate the methodology for hierarchical systems, let us assume that various networks labeled with numbers $a=1,\,\dots,\,m$ are treated as nodes in the system of coupled networks as it is presented at Fig. \ref{fig:fig4}. Here Eq.   (\ref{eq:NNON}) describes the transport of particles between networks and has the same structure as Eq. (\ref{eq:dynamics}) that describes the transport of particles between individual nodes. The equivalence between lower and higher level dynamics exists when parameters $\hat{f}_a$, $\hat{g}_a$ and $\hat{A}_{ab}$, $a,\,b=1,\,\dots,\,m$ for a higher level ($\hat{X}$ stands for $X$ at a higher level) are appropriately defined. This can be done in several ways, e.g.: 
	\begin{align}
	\hat{f}_a=M^{(a)}f^{(a)}_\mathrm{av},\;
	\hat{A}_{ab}=p_c^{(a\,b)},\;\;
	\hat{A}_{aa}=\frac{(f^{(a)}_ig^{(a)}_i)_{\mathrm{av}}\alpha^{(a\,a)}}{M^{(a)}\left(f_\mathrm{av}^{(a)}\right)^2}.\label{eq:levels}
	\end{align}
	It follows that one can extrapolate iterations of our approach to the next level where networks are grouped into several network clusters and the density of interlinks between two networks belonging to the same cluster is much higher than  the corresponding density between two networks from different clusters. One can then imagine clusters of clusters and consider a   nested system of hierarchically built weakly-coupled networks.
	
	Eqs. (\ref{eq:levels}) after short algebra lead to  
	$\hat{g}_a=\sum_l \hat{A}_{la}\hat{f}_a=(f^{(a)}_ig^{(a)}_i)_{\mathrm{av}}/f_\mathrm{av}^{(a)}$. Thus the ergodic distributions of particles in different sub-networks can be expressed as  
	\begin{equation}
	\frac{N^{(a)}_\infty}{N}=\frac{M^{(a)}(f^{(a)}_ig^{(a)}_i)_{\mathrm{av}}}{\sum_b M^{(b)}(f^{(b)}_ig^{(b)}_i)_{\mathrm{av}}},
	\label{Nainfty}
	\end{equation}
	and the ergodic distribution of particles at   individual nodes is $n_i^{(a)}=N^{(a)}_\infty \mu_i^{(a)} =Nf_i^{(a)}g_i^{(a)}/\sum_{b=1}^m \sum_{j=1}^{M^{(b)}}f_j^{(b)} g_j^{(b)}$. Let us note that neither $N^{(a)}_\infty$ nor $\mu_i^{(a)}$  depends on inter-network connectivity matrix $p_c^{(a\,b)}$ that settles inter-networks diffusion constants $D^{(ab)}$ and  corresponding forces  $F^{(ab)}$. The matrix $p_c^{(ab)}$ sets rates the equilibrium  distribution of particles at different sub-networks is reached but it does not influence equilibrium densities.         
	
	\section{Conclusions} \label{SecVIII}
	
	In conclusion, we have developed a framework for analytical studies of diffusion processes on weakly coupled networks and nested hierarchical systems where links are sparser at higher levels than at lower ones. The approach is based on the time scales separation and the adiabatic approximation for particles distribution.  The approach  allows to calculate effective diffusion constants and driving forces at different levels within the system. The driving forces depend not only on  fitness factors of individual nodes but also on the topological features of networks i.e.  link densities. In this way, the flow of particles can be induced and controlled by applying appropriate changes to the system's structure. Entropy changes result from the transfers of individual particle entropy between different networks as well as from the entropy at the macroscopic level, which in turn relates to the lack of information about the total number of particles contained in a given sub-network. The equilibrium distribution of particles between different sub-networks depends only on  attractiveness of their  nodes and their local neighbors. 
	We have shown  that if 
	communities are sparsely  connected then  diffusion flows concentrate only in communities'
	interiors and the inter-network diffusion constant decreases  while
	decreasing the interlinks density and increasing  density of links inside the communities.

	This new approach can be also applied to the reverse engineering of hidden nested network structures by the observation of the diffusion processes \cite{simonsen,Shandilya,Autariello,Palla,Wang,Shen,Comin,Pinto,Brockmann,Cheng}. In fact, if several characteristic diffusion time scales are detected then one can presume the system contains nested hierarchical structures \cite{hierach,Rosvall_Hierarch,Ravasz,Sales-Pardo,Clauset}.

	\section*{Acknowledgements} 
	
	The work was supported by EU FP7 projects {\it DynaNets}, grant No 233847. JAH was also supported by   {\it Sophocles}, grant No. 317534, as RENOIR Project by the European Union Horizon 2020 research and innovation programme under the Marie Sk{\l}odowska-Curie grant agreement  691152,  by Ministry of Science and Higher Education (Poland),
	grant Nos.  W34/H2020/2016, 329025/PnH /2016,   and National Science Centre, Poland, grant No. 2015/19/B/ST6/02612 a Grant from The Netherlands Institute for Advanced Study in the Humanities and Social Sciences (NIAS) and  by the Russian Scientific Foundation, Agreement \#17-71-30029 with co-financing of Bank Saint Petersburg.

	\appendix

	\section{Formalism extension to weighted networks of networks with self-loops}\label{sec:non}
	Our approach developed for two coupled networks as presented in the Secs. \ref{SecIII} and \ref{SecV} can be extended, as we remarked, to the case of any networks of networks with weighted links and self-loops that link a vertex to itself. To see that let us consider $m$ weighted networks with  symmetrical adjacency matrices $A^{(a)}\in\mathbb{M}^{M^{(a)}\times M^{(a)}}(\mathbb{R}_+)$, $a=1,\,\dots,\,m$. Every network is described by a vector of the fitness factors $\mathbf{f}^{(a)}\in\mathbb{R}_+^{M^{(a)}}$ and a vector of  neighborhood attractiveness of the non-connected network $\mathbf{g}^{(a)}\in\mathbb{R}_+^{M^{(a)}}$, $g^{(a)}_{l_a}=\sum_r A^{a}_{r{k_a}}f^{(a)}_r$, $l_a=1,\,\dots,\,M^{(a)}$, and $a=1,\,\dots,\,m$, so it is a natural generalization of the case considered in the main part of the paper. Networks are connected one to another, which is described by probabilities $p_c^{(a\,b)},\;a,\,b=1,\,\dots,\,m$, $a\neq b$ for the link existence between  nodes belonging to two different networks $a$ and $b$.   Number of particles $N^{(a)}(t)$ can be expressed using the above notation as 
	
	\begin{align}
	&N^{(a)}(t+1)=\sum_{l=1}^{M^{(a)}}n_l^{(a)}(t+1)=\sum_{l=1}^{M^{(a)}}\left(\sum_{l_1=1}^{M^{(1)}} \frac{f_{l}^{(a)}p_c^{(a\,1)}n_{l_1}^{(1)}(t)}{g^{(1)}_{l_1}+p_c^{(2\,1)}M^{(2)}f_{\mathrm{av}}^{(2)}+\dots
		+p_c^{(m\,1)}M^{(m)}f_{\mathrm{av}}^{(m)}}+\dots+\right.\nonumber\\ &+\left.\sum_{l_a=1}^{M^{(a)}}\frac{f^{(a)}_{l}A^{(a)}_{ll_a}n_{l_a}^{(a)}(t)}{p_c^{(1\,a)}M^{(1)}f_{\mathrm{av}}^{(1)}+\dots
		+g^{(a)}_{l_a}+\dots+p_c^{(m\,a)}f_{l}^{(m)}M^{(m)}}+
	\dots+\sum_{l_m=1}^{M^{(m)}}\frac{f_{l}^{(a)}p_c^{(a\,m)}n_{l_m}^{(m)}(t)}{p_c^{(1\,m)}M^{(1)}f_{\mathrm{av}}^{(1)}
		+\dots+g^{(m)}_{l_m}} \right)=\nonumber\\
	&=\sum_{l_1=1}^{M^{(1)}} \frac{f_{\mathrm{av}}^{(a)}M^{(a)}p_c^{(a\,1)}n_{l_1}^{(1)}(t)}{g^{(1)}_{l_1}+p_c^{(2\,1)}M^{(2)}f_{\mathrm{av}}^{(2)}+\dots
		+p_c^{(m\,1)}M^{(m)}f_{\mathrm{av}}^{(m)}}+\dots+ \sum_{l_a=1}^{M^{(a)}}\frac{g^{(a)}_{l_a}n_{l_a}^{(a)}(t)}{p_c^{(1\,a)}M^{(1)}f_{\mathrm{av}}^{(1)}+\dots
		+g^{(a)}_{l_a}+\dots+p_c^{(m\,a)}f_{l}^{(m)}M^{(m)}}+\nonumber\\&\dots+\sum_{l_m=1}^{M^{(m)}}\frac{p_c^{(a\,m)}f_{\mathrm{av}}^{(a)}M^{(a)}n_{l_m}^{(m)}(t)}{p_c^{(1\,m)}M^{(1)}f_{\mathrm{av}}^{(1)}
		+\dots+g^{(m)}_{l_m}}.\label{eq:calculation}
	\end{align}
	We assume that a density of internal network  connections is much larger than  a density  of inter-network connections, similarly as it was for the two-networks case. This assumption leads to the  following inequality
	\begin{equation}\label{eq:assumption}
	g_{l_a}^{(a)}\gg\sum_{b\neq a}^m p_c^{(a\,b)}M^{(b)}f_{\mathrm{av}}^{(b)},
	\end{equation}
	where $a=1,\,\dots,\,m$, $l_a=1,\,\dots,\,M^{(a)}$, 
	$q\neq a$. Similarly to the two-networks case the assumption (\ref{eq:assumption}) implies separation of time scales i.e. the diffusion inside  every network is much faster than the diffusion between the networks. This fact allows to write the number of particles in the $l_a$-th node  in the $a$-th network at time step $t$ in the adiabatic approximation as 
	\begin{equation}\label{eq:density}
	n^{(a)}_{l_a}(t)=N^{(a)}(t)\mu_{l_a}^{(a)}.
	\end{equation}
	
	Using Eq. (\ref{eq:assumption}) after some algebra one can write the rhs of Eq. (\ref{eq:calculation}) as    
	\begin{align*}
	&N^{(a)}(t+1)=\\&=\sum_{l_1=1}^{M^{(1)}} \frac{f_{\mathrm{av}}^{(a)}M^{(a)}p_c^{(a\,1)}}{g^{(1)}_{l_1}}n_{l_1}^{(1)}(t)+\dots+ \sum_{l_a=1}^{M^{(a)}}\left((1-\frac{\sum_{r\neq a}f_{\mathrm{av}}^{(r)}M^{(r)}p_c^{(r\,a)}}{g^{(a)}_{l_a}}\right)n_{l_a}^{(a)}(t)+\dots+
	\sum_{l_m=1}^{M^{(m)}}\frac{f_{\mathrm{av}}^{(a)}M^{(a)}p_c^{(a\,m)}}{g^{(m)}_{l_m}}n_{l_m}^{(m)}(t).
	\end{align*}
	
	Since Eq. (\ref{eq:n_i}) is still valid for weighted networks with self-loops thus Eq. (\ref{eq:density}) combined with the formula for the equilibrium density for a single  network   leads us to the equation for particles flows in a system of network of networks  
	\begin{equation}
	N^{(a)}(t+1)=\sum_{l=1}^{m}\alpha^{(a\,l)}N^{(l)}(t),\;\;\mathrm{where}\;\;\alpha^{(a\,b)}=p_c^{(a\,b)}M^{(a)}\frac{f^{(a)}_{\mathrm{av}}f^{(b)}_{\mathrm{av}}}{(f^{(b)}_ig^{(b)}_i)_{\mathrm{av}}},\;\alpha^{(a\,a)}=1-\sum_{r\neq a}\alpha^{(r\,a)},
	\end{equation}
	which completes our calculations for the nested systems.\\

	
	
	
	
\end{document}